\documentclass[preprint,amsmath,amssymb,aps,showkeys,showpacs]{revtex4}
\usepackage[english]{babel}
\usepackage{graphicx}
\usepackage{graphics}
\usepackage{amsmath}
\usepackage{dcolumn}
\usepackage{amssymb}
\usepackage{bm}

\begin{document}
\title{Fermions in G\"odel-type background space-times with torsion and the Landau quantization}
\author{G. Q. Garcia}
\email{gqgarcia@fisica.ufpb.br}
\affiliation{Departamento de F\'isica, Universidade Federal da Para\'iba, Caixa Postal 5008, 58051-900, Jo\~ao Pessoa-PB, Brazil.}

\author{J. R. de S. Oliveira}
\email{jardson.ricardo@gmail.com}
\affiliation{Departamento de F\'isica, Universidade Federal da Para\'iba, Caixa Postal 5008, 58051-900, Jo\~ao Pessoa-PB, Brazil.} 

\author{K. Bakke}
\email{kbakke@fisica.ufpb.br}
\affiliation{Departamento de F\'isica, Universidade Federal da Para\'iba, Caixa Postal 5008, 58051-900, Jo\~ao Pessoa-PB, Brazil.}

\author{C. Furtado}
\email{furtado@fisica.ufpb.br} 
\affiliation{Departamento de F\'isica, Universidade Federal da Para\'iba, Caixa Postal 5008, 58051-900, Jo\~ao Pessoa-PB, Brazil.}

\begin{abstract}
In this paper, we analyze Dirac fermions in   G\"odel-type background space-times with torsion. We also consider the G\"odel-type spacetimes  embedded in a topological defect background. We show that relativistic bound states solutions to the Dirac equation can be obtained by dealing with three cases of the G\"odel-type solutions with torsion, where a cosmic string  passes through these three cases of  the space-time. We obtain the relativistic energy levels for all cases of the G\"odel-type solutions with torsion with a cosmic string, where we show that there exists an analogy with the Landau levels for Dirac particles. We also show that the presence of torsion in the space-time yields new contributions to the relativistic spectrum of energies and  that the presence of the topological defect modifies the degeneracy of the relativistic energy levels.  

\end{abstract}


\maketitle

\section{Introduction}\label{sec1}

In 1949, G\"odel \cite{godel} obtained the first cosmological solution to Einstein equations for  the rotating matter. This stationary solution to the equations of general relativity was obtained by considering a spatially homogeneous  space-time with cylindrical symmetry. This solution has interesting physical properties,  with the first of them is the possibility of closed timelike curves (CTCs). Hawking \cite{hawking} studied the presence of CTCs and conjectured that  their presence is physically inconsistent. Rebou\c cas {\it et al} \cite{reboucas,reboucas2,reboucas3} went further and generalized this metric in cylindrical  coordinates. They  considered in detail the problem of causality, and then, established three classes of solutions that are characterized by the following properties: (i) solutions where there are no CTCs; (ii) solutions where there  is  a sequence of  alternating causal and non-causal regions; finally, the case (III) where the solutions are characterized by only one non-causal region. Besides, Dabrowski \cite{dabrowski}   established some criteria in order to define the possibility of the existence of CTCs,  where  the quantities called super energy and super-momentum  were introduced for this purpose. In the context  of the  String-Theory \cite{bertolami},  the appearance of CTCs in  the G\"odel space-time was investigated in Ref. \cite{barrow}. Several properties of the G\"odel solution  were analyzed by  Barrow \cite{barrow2}.

The interest in the study of the problems  involving G\"odel-type solutions has attracted attention of several authors.  Recently, the study of the properties of these kinds of space-times have been investigated in the light of the equivalence problem techniques in  the Riemannian G\"odel-type space-times \cite{reboucas}  and for Riemann-Cartan G\"odel-type space-times \cite{joeljmp,joel,joelrebou}. The G\"odel-type solution has a great importance and  it has motivated a large number of studies,  for instance,  of rotating G\"odel-type models as well as  of causal anomalies in the framework of other theories of gravitation: the hybrid metric-Palatini gravity \cite{reboupala}, the Chern-Simons modified gravity theory \cite{furgodel,furgodel3}, the Horava-Lifshitz theory of gravity \cite{furgodel2,joelhorava} and the Brans-Dick theory of gravity \cite{porfi}. The family of G\"odel-type solutions has motivated a considerable number of investigations  of the geodesics  in G\"odel-type solutions in several physical  contexts, for example, see Refs. \cite{figueiredo,fiol,calvao,camci,buser}. Recently, the spherical G\"odel-type solution has been used to  investigate electronic properties in a geometric  model for describing the fullerene molecule \cite{everton}. The study of the quantum dynamics of scalar particles and spin-1/2 particles have been investigated by several authors in G\"odel-type space-times. The first  study addressing this problem was  carried out by Figueiredo {\it et al} \cite{figueiredo}, where the Klein-Gordon and Dirac equations in the G\"odel-type space-time with positive and negative curvatures, and in the flat G\"odel-type space-time  were studied. Drukker {\it et al} \cite{fiol} examined the relationship between the quantum dynamics of a scalar particle in the four-dimensional class of G\"odel solutions in general relativity with Landau levels in the flat, spherical and hyperbolic spaces. They investigated the similarity between the energy levels of the scalar quantum particle moving in a class of G\"odel-type space-times,  where the eigenvalues of the Landau problem are obtained in backgrounds with  zero, positive and negative curvatures. Guided by the similarity with the Landau problem, they also studied the possible holographic description of a single chronologically safe region of this class of space-times. This analogy was also observed by Das and Gegenberg \cite{gegenberg} by studying the quantum dynamics of scalar particles in the Som-Raychaudhuri space-time (G\"odel flat solution) and compared with the Landau levels in the flat space. Recently, one of us \cite{josevi} have investigated the Klein-Gordon equation in class of the G\"odel solutions with a cosmic string passing through the  space-time. It was demonstrated that the presence of cosmic string break the degeneracy of relativistic energy levels. The quantum dynamic  of a scalar particle in the Som-Raychaudhuri space-time have also been investigated in Ref. \cite{epjpchina}. In recent years, the study of the quantum dynamics of a particle in the presence of  an uniform magnetic field has been investigated in different types of the curved spaces. It is worth mentioning the studies of the Landau problem in hyperbolic space \cite{comtet,dunne}, in the spherical space \cite{dunne} and in the presence of topological defects \cite{prdknut}. 

 Basing on the wide interest  to the study of problems related  to G\"odel-type solutions, we investigate the quantum dynamics of spin-$1/2$ particles in this family of   background space-times with torsion in the presence of a topological defect.   We  will study the influence of torsion and the cosmic string through Einstein-Cartan theory. We solve the Dirac equation in this set of G\"odel solutions. Recently, a series of studies have been made with the purpose of investigating the quantum and classical dynamics of particles in curved spaces with topological defects and  the possible detection of this defect. We claim that the studies of these problems in the present paper can be useful to investigate the influence of the cosmic string in the G\"odel-type  background space-times  with torsion. The techniques used in this article can be  applied to study the influence of topological defects in condensed matter systems as well as to investigate  the Hall effect in curved surfaces with rotation and with the presence of topological defects. In this paper, we investigate the relativistic quantum dynamics of a Dirac particle in  the presence of a topological defect in a class of G\"odel  space-times with torsion. We solve the Dirac equation in Som-Raychaudhury, spherical and hyperbolic background space-times with torsion  and a cosmic string  passing through them. We obtain the eigenvalues of energy in all three cases and observe the similarity with Landau levels for a spin-$1/2$ particle. We also observe that presence of the topological defect breaks the degeneracy of the relativistic energy levels, and the eigenfunctions depend on the parameter  characterizing the presence of the topological defect in  these background space-times  with torsion.  We also discuss the influence of torsion of the family of G\"odel-type solutions on the energy levels. 

This contribution is organized as follows: in section \ref{secgod}, we present the G\"odel-type solution with torsion and a cosmic string passing through the background  space-times; in section \ref{sec2}, the Dirac equation in the background space-times  with torsion is written for the flat, spherical and hyperbolic cases; in section \ref{sec3}, relativistic bound state solutions to the Dirac equation in G\"odel-type background space-times  with torsion are investigated; finally, in section \ref{sec4}, we present the conclusions. In this paper, we use natural units $\left(\hbar=c=G=1\right)$.

\section{Backgrounds of the G\"odel-Type space-time with a topological defect}\label{secgod}

 In this section, we make a brief review of a class of G\"odel-type solutions with a topological defect. In a very interesting paper, Rebou\c cas and Tiomno \cite{reboucas} obtained a generalization of the original  G\"odel metric, where the solution is characterized by a vorticity $\Omega$. They made a detailed and extensive discussion about the homogeneity and isotropy of this Riemannian manifold by showing that it is characterized by two parameters, the vorticity $\Omega$ and the parameter $l$ associated with the curvature radius. They also obtained all homogeneous solutions with a rigid rotation in the general relativity context. These  classes of background space-times  lead to new solutions  of the Einstein field equations which can be called as the G\"odel-type solutions. Recently, several authors \cite{sandro,josevi} have studied   generalizations of G\"odel-type solutions for cosmic string space-times. In this work, we are interested in the study of the quantum dynamics of relativistic fermions in  this class of space-times. Due to the cylindrical symmetry imposed by the defect, we consider the space-time metric  representing a class of the G\"odel-type solution with torsion and a topological defect to be written in cylindrical coordinates $(t,r,\phi,z)$: 
\begin{eqnarray}
\label{geoplain}
ds^{2}=-\left(dt+\alpha\Omega\frac{\sinh^{2}lr}{l^{2}}\,d\phi\right)^{2}+\alpha^{2}\frac{\sinh^{2}2lr}{4l^{2}}\,d\phi^{2}+dr^{2}+dz^{2}.
\end{eqnarray}
The coordinates $(t,\,r,\,\phi,\,z)$ are defined in the ranges: $0\leq r <\infty$, $0\leq\phi\leq 2\pi$ and $-\infty<(z,t)<\infty$. Moreover, the parameter $\Omega$ characterizes the vorticity of the space-time, and the parameter $\alpha$ characterizes the cosmic string, since it is associated with the deficit of angle $\alpha=(1-4\lambda)$, with $\lambda$  being the mass per unit length of the cosmic string, and it assumes values in the range $0<\alpha<1$. Note that Eq. (\ref{geoplain}) represents a family of G\"odel-type solutions with the presence of a cosmic string. We can obtain this solution in  the general relativity or,  for the non-zero torsion,  in the Einstein-Cartan theory. For the case $\alpha=1$, Rebou\c cas and Tiomno \cite{reboucas} investigated the  G\"odel-type of Riemannian space-time and studied the possible sources of the model represented by (\ref{geoplain}) in  general relativity. If  we restrict the matter content to a perfect fluid, an electromagnetic field and a massless scalar field, we can only yield  the line element with $-\infty<{l^{2}}<\Omega^{2}$ as solutions  for the appropriate coupled field equations (Einstein-Maxwell-Klein-Gordon equations). In Ref. \cite{oliveira}, Oliveira {\it et al}  obtained that it is possible to generate in the Einstein-Cartan theory all line elements $-\infty<{l^{2}}<\infty$ by taking into account only a Weyssenhoff-Raabe perfect fluid \cite{weyss} for  the matter content,  since,  in the context of the Einstein-Cartan theory with a torsion, only this non-vanishing component corresponds to a Weyssenhoff fluid,  whose vector associated to the spin density is aligned along the direction of the rotation vector (the $z$-axis) \cite{oliveira}. Throughout this article, we shall refer to this torsion as polarized (aligned) along the rotation vector. Clearly, this torsion also displays the  same translational symmetry of the metric (\ref{geoplain}). The authors of Refs. \cite{joelrebou,joel} denominated this class of family of space-time manifolds as the Riemann-Cartan G\"odel-type manifolds. In this way, the solution (\ref{geoplain}) represents the family of Riemann-Cartan G\"odel-type manifolds with the presence of a topological defect. Now, let us present the condition for the existence of closed time-like curves (CTC) in the line element given in Eq. (\ref{geoplain}):
\begin{eqnarray}\label{CTCcon}
\tanh\left(l\,r_{c}\right)=\frac{l}{\Omega}.
\end{eqnarray}
The parameter $r_{c}$ is a critical radius:  only in the region outside of it ($r>r_{c}$), the CTCs can exist. On the other hand, for $r<r_{c}$, the CTCs cannot exist and this region is known as the chronologically safe region. From Eq. (\ref{CTCcon}), we can observe that the presence of the cosmic string does not affect the condition of existence of CTC since the expression (\ref{CTCcon}) is independent of parameter $\alpha$. Therefore, the condition for the existence of CTCs in Eq. (\ref{CTCcon}) is the  same that obtained for the case without the presence of a topological defect. Some interesting limits of the metric (\ref{geoplain}) are: I) for $l\rightarrow 0$, we obtain the geometry of the Som-Raychaudhuri space-time \cite{som} with a cosmic string; II) for $\Omega=0$ and $\alpha=1$, we have the Minkowski space-time limit; III) for $l^{2}=\Omega^{2}/2$ and $\alpha=1$, we obtain the original solution obtained by G\"odel \cite{godel}; IV) the limits where $l^{2}=\Omega^{2}$ and $\alpha=1$, we obtain the anti-de Sitter space-time. For the particular case where $\Omega=0$ and $\alpha \neq1$, we obtain the line element:
\begin{eqnarray}\label{string}
ds^{2}=-dt^{2} + dr^{2}+\alpha^{2}r^{2}d\phi^{2}+ dz^{2},
\end{eqnarray}
which represents the cosmic string space-time \cite{Vil1,His}, one of the most important examples of the topological defects. Several authors have dealt with curved space-times with the presence cosmic strings by investigating the influence of this topological defect on physical systems.  The Schwarzschild space-time with cosmic string was studied by Mukunda \cite{mukunda} and Germano and Bezerra \cite{germano}, where the classical and quantum  behavior of particles demonstrated the influence of topological defects on the states of the particle.  Gal'tsov and Masar \cite{22} studied the Kerr space-time with cosmic string. The field theory in AdS space-time with a cosmic string was investigated by de Mello and Saharian \cite{demellosahar}. In a recent work, Fernandes {\it et al} \cite{sandro} have solved the Klein-Gordon equation for a particle confined to two concentric thin shells in the G\"odel, Kerr-Newman and FRW space-times with the presence of topological defect passing through then.

\section{Dirac Equation in curved  background space-times with torsion}\label{sec2}

In this section, we start by making a brief review of the Dirac equation in {\bf a} curved space-time with the presence of torsion. Following  the spinor theory in curved space \cite{naka,weinberg,cartan1}, the Dirac equation must be written in terms of the components of the covariant derivative $\nabla_{\mu}=\partial_{\mu}+ \Gamma_{\mu}\left(x\right)$, where $\Gamma_{\mu}\left(x\right)=\frac{i}{4}\omega_{\mu ab}\left(x\right)\Sigma^{ab}$ is called as the spinorial connection \cite{bd,naka}. However,  in the presence of torsion, the components of the covariant derivative  become \cite{shap,shap2}:
\begin{eqnarray}
\nabla_{\mu}=\partial_{\mu}+\frac{i}{4}\omega_{\mu ab}\left(x\right)\Sigma^{ab}+\frac{i}{4}K_{\mu ab}\left(x\right)\Sigma^{ab},
\label{2.1}
\end{eqnarray}
where $\Sigma^{ab}=\frac{i}{2}\left[\gamma^{a},\,\gamma^{b}\right]$ and the $\gamma^{a}$ matrices are the Dirac matrices defined in the Minkowski space-time \cite{bd,greiner}. The connection $K_{\mu ab}\left(x\right)$ is related with the contortion tensor $K_{\beta,\nu\,\mu}\left(x\right)$, in turn, it is related to the torsion tensor $T_{\beta\nu\mu}\left(x\right)$. By following Refs. \cite{shap,shap2}, we can write the torsion tensor in terms of three irreducible components: the trace 4-vector $T_{\mu}\left(x\right)=T^{\beta}_{\,\,\,\mu\beta}\left(x\right)$, the axial 4-vector $S^{\nu}\left(x\right)=\epsilon^{\nu\alpha\beta\mu}\,T_{\alpha\beta\mu}\left(x\right)$, and the tensor $q_{\alpha\beta\mu}\left(x\right)$ satisfying the following relations: $q^{\beta}_{\,\,\,\mu\beta}=0$ and $\epsilon^{\alpha\beta\mu\nu}\,q_{\beta\mu\nu}=0$. Thereby, after some calculation, we can write the Dirac equation in the form \cite{shap,shap2,bf}:
\begin{eqnarray}
i\gamma^{\mu}\,\partial_{\mu}\,\psi+i\gamma^{\mu}\,\Gamma_{\mu}\left(x\right)\psi-\frac{1}{8}\,\gamma^{\mu}\gamma^{5}\,S_{\mu}\left(x\right)\,\psi=0.
\label{2.2}
\end{eqnarray}
where $\gamma^{\mu}=e^{\mu}_{\,\,\,a}\left(x\right)\,\gamma^{a}$ and $\gamma^{5}=i\gamma^{0}\,\gamma^{1}\,\gamma^{2}\gamma^{3}$. Note that the Dirac equation above does not depend on $T_{\mu}\left(x\right)$ and $q_{\alpha\beta\mu}\left(x\right)$. This occurs since these terms do not couple with fermions, but the axial 4-vector ($S^{\mu}$)  couples with fermions as shown in Refs. \cite{shap,shap2}.

In the next subsections, we write the Dirac equation in the space-time background determined by the line element (\ref{geoplain}) for three particular cases: for $l=0$, $l^{2}<0$ and $l^{2}>0$, respectively.

\subsection{Dirac equation in a Som-Raychaudhuri space-time}\label{subsec1}

Let us consider $l=0$ in the line element (\ref{geoplain}), then, the G\"odel-type space-time  reduces to the Som-Raychaudhuri solution \cite{som} with the presence of a cosmic string. Thereby, the line element becomes 
\begin{eqnarray}
ds^2 = -\left(dt+\alpha\,\Omega\,r^{2}\,d\phi\right)^{2}+dr^{2} + \alpha^{2}\,r^{2}\,d\phi^{2}+dz^{2}. 
\label{2.A.1}
\end{eqnarray}

In order to work with fermions in this scenario, let us define the components of the non-coordinate basis $\hat{\theta}^{a}=e^{a}_{\,\,\,\mu}\left(x\right)\,dx^{\mu}$ and its inverse $dx^{\mu}=e^{\mu}_{\,\,\,a}\left(x\right)\hat{\theta}^{a}$ in the following form:
\begin{eqnarray}
e^{a}_{\,\,\,\mu}\left(x\right)= \left(
\begin{array}{cccc}
 1 & 0 & \alpha\,\Omega\,r^{2} & 0\\ 
0 & 1 & 0 & 0\\
 0 & 0 & \alpha\,r & 0\\
 0 & 0 & 0 & 1\\ 
\end{array}\right);\,\,\,\,\,
e^{\mu}_{\,\,\,a}\left(x\right)=\left(
\begin{array}{cccc}
 1 & 0 & -\Omega\,r & 0\\ 
0 & 1 & 0 & 0\\
 0 & 0 & \frac{1}{\alpha\,r} & 0\\
 0 & 0 & 0 & 1\\ 
\end{array}\right).
\label{2.A.2}
\end{eqnarray}

By solving the Maurer-Cartan structure equations $T^{a}=d\hat{\theta}^{a}+\omega^{\,\,\,a}_{\mu\,\,\,b}\left(x\right)\,dx^{\mu}\wedge\hat{\theta}^{b}$, where $T^{a}=\frac{1}{2}\,T^{a}_{\,\,\,\mu\nu}\left(x\right)\,dx^{\mu}\wedge\,dx^{\nu}$ is the torsion 2-form, $\omega^{\,\,\,a}_{\mu\,\,\,b}\left(x\right)$ is the spin connection, $T^{a}_{\,\,\,\mu\nu}\left(x\right)=e^{a}_{\,\,\,\beta}\left(x\right)\,T^{\beta}_{\,\,\,\mu\nu}\left(x\right)$, $d$ is  the exterior derivative and the symbol $\wedge$ is the wedge product \cite{cartan2,naka,carroll}, we obtain two non- zero components of the spin connection and the torsion tensor:
\begin{eqnarray}
\omega^{\,\,\,1}_{\phi\,\,\,2}\left(x\right)=-\omega^{\,\,\,2}_{\phi\,\,\,1}\left(x\right)= -\alpha;\,\,\,\,T^{t}_{\,\,\,r\,\phi}=-T^{t}_{\,\,\,\phi\,r}=2\alpha\,\Omega\,r.
\label{2.A.3}
\end{eqnarray}

Since we have written the Dirac equation (\ref{2.2}) in terms of the irreducible components of the torsion tensor, then, from Eq. (\ref{2.A.3}), we obtain only one non-zero component of the axial 4-vector,  that  is
\begin{eqnarray}
S^{z}=-2\Omega.
\label{2.A.7}
\end{eqnarray} 

Therefore, by substituting Eqs. (\ref{2.A.3}) and (\ref{2.A.7}) into Eq. (\ref{2.2}), we have
\begin{eqnarray}
i\gamma^{0}\frac{\partial\psi}{\partial t}+i\gamma^{1}\left(\frac{\partial}{\partial r}+\frac{1}{2r}\right)\psi+\frac{i\gamma^{2}}{\alpha\,r}\left(\frac{\partial}{\partial \phi}-\alpha\,\Omega\,r^{2}\,\frac{\partial}{\partial t}\right)\psi+ i\gamma^{3}\left(\frac{\partial}{\partial z}+i\frac{S^{z}}{8}\gamma^{5}\right)\psi=M\psi,
\label{2.A.9}
\end{eqnarray}
where the $\gamma^{a}$ matrices are \cite{greiner} 
\begin{eqnarray}
\gamma^{0}=\left(
\begin{array}{cc}
1 & 0 \\
0 & -1 \\
\end{array}\right);\,\,\,\,\,\,
\gamma^{k}=\left(
\begin{array}{cc}
 0 & \sigma^{k} \\
-\sigma^{k} & 0 \\
\end{array}\right),
\label{2.8}
\end{eqnarray}

Note that  Eq. (\ref{2.A.9}) involves the parameters $\Omega$ and $\alpha$ characterizing  the rotation and the topological defect, respectively. Hence, Eq. (\ref{2.A.9}) is the Dirac equation in the Som-Raychaudhuri space-time with a cosmic string.

\subsection{Dirac equation in the spherical symmetrical G\"odel-type space-time}\label{subsec2}

Let us consider the case $l^{2}<0$ in the line element (\ref{geoplain}), which corresponds to the G\"odel-type space-time with a spherical symmetry in the presence of a cosmic string.  In this case, we introduce new coordinates $R= i/2l$ and $\theta=r/R$ \cite{josevi}, then, the line element (\ref{geoplain}) becomes
\begin{eqnarray}
ds^{2}=-\left(dt+\frac{\alpha\,\Omega\,r^{2}}{1+r^{2}/4R^{2}} d\phi\right)^2 +\left(1+\frac{r^{2}}{4R^{2}}\right)^{-2} \left(dr^{2}+\alpha^{2}\,r^{2}\,d\phi^{2}\right)+dz^{2}, 
\label{2.B.1}
\end{eqnarray}
and we can define
\begin{eqnarray}
e^{a}_{\,\,\,\mu}\left(x\right)=\left(
\begin{array}{cccc}
1 & 0 & \frac{\alpha\,\Omega\,r^{2}}{1 + r^{2}/4R^{2}} & 0\\
0 & \left(1 + \frac{r^{2}}{4R^{2}}\right)^{-1} & 0 & 0\\ 
0 & 0 & \frac{\alpha r}{1 + r^{2}/4R^{2}} & 0\\
 0 & 0 & 0 & 1\\ 
\end{array}\right);
e^{\mu}_{\,\,\,a}\left(x\right)= \left(
\begin{array}{cccc}
 1 & 0 & -\Omega\,r & 0\\
 0 & \left(1+\frac{r^{2}}{4R^{2}}\right) & 0 & 0\\
 0 & 0 & \frac{\left(1 + r^{2}/4R^{2}\right)}{\alpha\,r} & 0\\
 0 & 0 & 0 & 1\\ 
\end{array}\right).
\label{2.B.2}
\end{eqnarray} 

By solving the Maurer-Cartan structure equations \cite{naka}, we obtain
\begin{eqnarray}
\omega^{\,\,\,1}_{\phi\,\,\,2}\left(x\right)&=&-\omega^{\,\,\,2}_{\phi\,\,\,1}\left(x\right)= -\alpha\frac{\left(4R^{2}-r^{2}\right)}{\left(4R^{2}+r^{2}\right)};\nonumber\\
[-2mm]\label{2.B.3}\\[-2mm]
T^{t}_{\,\,\,r\,\phi}&=&-T^{t}_{\,\,\,\phi\,r}=\frac{2\,\alpha\,\Omega\,r}{\left(1+r^{2}/4R^{2}\right)^{2}}.\nonumber
\end{eqnarray}

From Eq. (\ref{2.B.3}), we also obtain just one non-zero component of the axial 4-vector, which is given by Eq. (\ref{2.A.7}). Thereby, Eq. (\ref{2.2}) becomes
\begin{eqnarray}
i\gamma^{0}\frac{\partial\psi}{\partial t}&+&i\gamma^{1}\left[\left(1+\frac{r^{2}}{4R^{2}}\right)\frac{\partial}{\partial r} +\left(1-\frac{r^{2}}{4R^{2}}\right)\frac{1}{2r}\right]\psi+\frac{i\gamma^{2}}{\alpha\,r}\left(1+\frac{r^{2}}{4R^{2}}\right)\frac{\partial\psi}{\partial \phi}\nonumber\\
[-2mm]\label{2.B.6}\\[-2mm]
&-&i\,\Omega\,r\,\gamma^{2}\,\frac{\partial\psi}{\partial t}+i\gamma^{3}\left(\frac{\partial}{\partial z}+i\frac{S^{z}}{8}\gamma^{5}\right)\psi=M\psi,\nonumber
\end{eqnarray} 
where the $\gamma^{a}$ matrices have been defined in Eq. (\ref{2.8}).

Again, we can observe a dependence of the Dirac equation on the parameters $\Omega$ and $\alpha$  characterizing the rotation and the topological defect, respectively. However, we deal with a space-time with  a positive curvature in Eq. (\ref{2.B.6}). Thereby, Eq. (\ref{2.B.6}) corresponds to the Dirac equation in the G\"odel-type space-time of  a positive curvature with the cosmic string.

\subsection{Dirac Equation in a hyperbolic G\"odel-type space-time}\label{subsec3}

Finally, we consider the case $l^{2}>0$ in the line element (\ref{geoplain}), which corresponds to the hyperbolic G\"odel-type space-time with the cosmic string \cite{josevi}. In this case, the line element (\ref{geoplain}) becomes
\begin{eqnarray}
ds^{2}=-\left(dt+\frac{\alpha\,\Omega\,r^{2}}{1- l^{2}\,r^{2}}\,d\phi\right)^{2}+\left(1-l^{2}\,r^{2}\right)^{-2}\left(dr^{2}+\alpha^{2}\,r^{2}\,d\phi^{2}\right)+dz^{2}, 
\label{2.C.1}
\end{eqnarray}
and we can establish
\begin{eqnarray}
e^{a}_{\,\,\,\mu}\left(x\right)=\left(
\begin{array}{cccc}
 1 & 0 & \frac{\alpha\,\Omega\,r^{2}}{1-l^{2}\,r^{2}} & 0\\ 
0 & \left(1-l^{2}\,r^{2}\right)^{-1} & 0 & 0\\
 0 & 0 & \frac{\alpha\,r}{1-l^{2}\,r^{2}} & 0\\ 
0 & 0 & 0 & 1\\ 
\end{array}\right);\,\,\,
\ e^{\mu}_{\,\,\,a}\left(x\right)=\left(
\begin{array}{cccc}
 1 & 0 & -\Omega\,r & 0\\
 0 & \left(1-l^{2}\,r^{2}\right) & 0 & 0\\
 0 & 0 & \frac{\left(1-l^{2}\,r^{2}\right)}{\alpha\, r} & 0\\ 
0 & 0 & 0 & 1\\ 
\end{array}\right).
\label{2.C.2}
\end{eqnarray}

Again, we solve the Maurer-Cartan equations \cite{naka} and obtain
\begin{eqnarray}
\omega^{\,\,\,1}_{\phi\,\,\,2}\left(x\right)&=& -\omega^{\,\,\,2}_{\phi\,\,\,1}\left(x\right)= -\alpha\,\frac{\left(1+l^{2}\,r^{2}\right)}{\left(1-l^{2}\,r^{2}\right)};\nonumber\\ 
[-2mm]\label{2.C.3}\\[-2mm]
T^{t}_{\,\,\,r\,\phi}&=&-T^{t}_{\,\,\,\phi\,r}=\frac{2\alpha\,\Omega\,r}{\left(1-l^{2}\,r^{2}\right)^{2}}.\nonumber
\end{eqnarray}

From Eq. (\ref{2.C.3}), we obtain the same component of the axial 4-vector given in Eq. (\ref{2.A.7}); thus, Eq. (\ref{2.2}) becomes
\begin{eqnarray}
i\gamma^{0}\frac{\partial\psi}{\partial t}&+&i\gamma^{1}\left[\left(1-l^{2}\,r^{2}\right)\frac{\partial}{\partial r}+\left(1+l^{2}\,r^{2}\right)\frac{1}{2r}\right]\psi+\frac{i\gamma^{2}}{\alpha\,r}\left(1-l^{2}\,r^{2}\right)\,\frac{\partial\psi}{\partial\phi}\nonumber\\
[-2mm]\label{2.C.6}\\[-2mm]
&-&i\,\Omega\,r\,\gamma^{2}\,\frac{\partial\psi}{\partial t}+i\gamma^{3}\left(\frac{\partial}{\partial z}+i\frac{S^{z}}{8}\gamma^{5}\right)\psi=M\psi,\nonumber
\end{eqnarray} 
where the $\gamma^{a}$ matrices have also been defined in Eq. (\ref{2.8}). In this case, we have a space-time with  a negative curvature, hence, Eq. (\ref{2.C.6}) corresponds to the Dirac equation in the G\"odel space-time of  a negative curvature with a cosmic string.

\section{Solutions to the Dirac equation and Landau Quantization}\label{sec3}

 In recent papers \cite{fiol,gegenberg}, the analogy between classical and quantum dynamics in these three classes of space-time  backgrounds and the study  of the particle dynamics in the presence of an external magnetic field in space-times of constant curvature have been made. In particular, in Ref. \cite{fiol}  it was obtained that the geodesic curves in these  background space-times are circles and  their similarity with orbits of  electrons moving under the influence of an orthogonal magnetic field is demonstrated, which are the well-known Larmor orbits. After describing the Dirac equation in three particular cases of the G\"odel-type space-time in the previous section, in this section we focus on the search for relativistic bound state solutions to the Dirac equations (\ref{2.A.9}), (\ref{2.B.6}) and (\ref{2.C.6}). For this purpose, we write Eqs. (\ref{2.A.9}), (\ref{2.B.6}) and (\ref{2.C.6}) in the following form from now on:  
\begin{eqnarray}
i\frac{\partial\psi}{\partial t}=\left[\left(\alpha^{1}\hat{\pi}_{r}+\alpha^{2}\hat{\pi}_{\phi}\right)+\Sigma^{3}\left(\hat{C}+\frac{S^{z}}{8}\right)\right]\psi,
\label{3.1}
\end{eqnarray}
where $\alpha^{i}= \gamma^{0}\gamma^{i}$ ($i=1,2, 3$), $\Sigma^{3}= \gamma^{5}\gamma^{0}\gamma^{3}$ and the operator $\hat{C}$ is defined as
\begin{eqnarray}
\hat{C} = \gamma^5\hat{\pi}_{z}+M\,\gamma^{3}\gamma^{5},
\label{3.2}
\end{eqnarray}
with $\hat{\pi}_{z}=-i\frac{\partial}{\partial z}$. Finally, the operators $\hat{\pi}_{r}$ and $\hat{\pi}_{\phi}$ given in Eq. (\ref{3.1}) will be defined latter.  

Observe that Eqs. (\ref{2.A.9}), (\ref{2.B.6}) and (\ref{2.C.6}) do not depend on the $z$-coordinate and the $\phi$-coordinate, thus, the operators $\hat{\pi}_{z}$ and $\hat{J}_{z}=-i\partial_{\phi}$ \cite{schu} commute with the Hamiltonian operator $\hat{H}$ given in the right-hand side of Eq. (\ref{3.1}) for all of the three cases of G\"odel-type background space-times, with $k=\mathrm{const}$ as being the eigenvalue of $\hat{\pi}_{z}$. For this reason, it is possible to show that the operator $\hat{C}$ commutes with the Hamiltonian operator $\hat{H}$ given in the right-hand side of Eq. (\ref{3.1}) in all of the cases given in the previous section. Therefore, the operators $\hat{C}$ and $\hat{H}$ share the same eigenstate basis \cite{sakurai}, where the eigenvalues of $\hat{C}$ and $\hat{J}_{z}=-i\partial_{\phi}$ are given by
\begin{eqnarray}
\hat{C}\psi&=& \nu\sqrt{k^{2}+M^{2}}\psi;\nonumber\\
[-2mm]\label{3.3}\\[-2mm]
\hat{J}_{z}\psi&=&j\psi=\left(m+\frac{1}{2}\right)\psi,\nonumber 
\end{eqnarray}   
where $\nu=\pm1$ and $m=0,\pm1,\pm2,\ldots$. Note that the operator $\hat{C}$ is proportional to the projection of spin along $z$-axis \cite{figueiredo}. With the definitions given in Eq. (\ref{3.3}), hence, we can write the solution to the Dirac equation (\ref{3.1}) in the form:
\begin{eqnarray}
\psi\left(t,\,r,\,\phi,\,z\right)=D\left(\begin{array}{c} \eta(r) \\ \chi(r)\end{array}\right)e^{-i\left(\mathcal{E}t-j\,\phi-k\,z\right)}, 
\label{3.4}
\end{eqnarray}
where 
\begin{eqnarray}
D=\left(
\begin{array}{cccc}
 1 & 0 & 0 & 0\\
 0 & 0 & 1 & 0\\ 
0 & 1 & 0 & 0\\ 
0 & 0 & 0 & 1 \end{array}\right);\,\,\,
\eta=u_{1}\left(r\right)\left(
\begin{array}{c}
 1\\ \tau_{1}\end{array}\right);\,\,\,
\chi=u_{2}\left(r\right)\left(
\begin{array}{c}
 1\\ \tau_{2}
\end{array}\right),  
\label{3.5}
\end{eqnarray}
and
\begin{eqnarray}
\tau_{i}=\frac{\mp M +\nu\sqrt{k^{2}+M^{2}}}{k},\,\,\, i=1, 2.
\label{3.6}
\end{eqnarray}

Now, we are able to solve the Dirac equations (\ref{2.A.9}), (\ref{2.B.6}) and (\ref{2.C.6}).

\subsection{Som-Raychaudhuri space-time with a topological defect}\label{subsec4}

Let us search for relativistic bound state solutions to the Dirac equation (\ref{2.A.9}), which corresponds to the case $l=0$ of the line element (\ref{geoplain}). Comparing Eqs. (\ref{2.A.9}) and (\ref{3.1}),  we see that the corresponding operators $\hat{\pi}_{r}$ and $\hat{\pi}_{\phi}$  turn out to be given by
\begin{eqnarray}
\hat{\pi}_{r}=-i\left(\frac{\partial}{\partial r}+\frac{1}{2r}\right);\,\,\,\,\hat{\pi}_{\phi}=\frac{-i}{\alpha\,r}\left(\frac{\partial}{\partial\phi}- \alpha\,\Omega\,r^{2}\frac{\partial}{\partial t}\right).
\label{3.A.1}
\end{eqnarray}
In this way, by substituting the {\it ansatz} (\ref{3.4}) into the Dirac equation (\ref{2.A.9}), we have a set of four coupled differential equations: 
\begin{eqnarray}
\left[\mathcal{E}-\left(\nu\sqrt{k^{2}+M^{2}}+\frac{S^{z}}{8}\right)\right]u_{1}&=&-i\tau_{2}\left[\frac{\partial}{\partial r}+\frac{1}{2r}+\frac{j}{\alpha\,r}+\mathcal{E}\,\Omega\,r\right]u_{2};\label{3.A.2}\\
\left[\mathcal{E}+ \left(\nu\sqrt{k^{2}+M^{2}}+\frac{S^{z}}{8}\right)\right]u_{2}&=&-i\tau_{1}\left[\frac{\partial}{\partial r}+\frac{1}{2r}-\frac{j}{\alpha\,r}- \mathcal{E}\,\Omega\,r\right]u_{1}\label{3.A.3};\\
\tau_{1}\left[\mathcal{E}-\left(\nu\sqrt{k^{2}+M^{2}}+\frac{S^{z}}{8}\right)\right]u_{1}&=&-i\left[\frac{\partial}{\partial r}+\frac{1}{2r}+\frac{j}{\alpha\,r}+ \mathcal{E}\,\Omega\,r\right]u_{2}\label{3.A.4};\\
\tau_{2}\left[\mathcal{E}+\left(\nu\sqrt{k^{2}+M^{2}}+\frac{S^{z}}{8}\right)\right]u_{2}&=&-i\left[\frac{\partial}{\partial r}+\frac{1}{2r}- \frac{j}{\alpha\,r} - \mathcal{E}\,\Omega\,r\right]u_{1}.\label{3.A.5}
\end{eqnarray}

From Eqs. (\ref{3.A.2})-(\ref{3.A.5}), we obtain a second order differential equation for $u_{1}$ and $u_{2}$ given by
\begin{eqnarray}
\frac{d^2 u_{i}}{dr^{2}}+\frac{1}{r}\frac{d u_{i}}{dr}-\left[\mathcal{E}^{2}\,\Omega^{2}\,r^{2}+ \frac{\zeta^{2}_{i}}{r^{2}}-\beta_{i}\right]u_{i}=0,
\label{3.A.6}
\end{eqnarray}
where $i=1,2$ are the  indices of each spinor component. The parameters $\zeta_{i}$ are defined as
\begin{eqnarray}
\zeta^{2}_{1}=\left(\frac{j}{\alpha} - \frac{1}{2}\right)^2,\ \ \zeta^{2}_{2} = \left(\frac{j}{\alpha} + \frac{1}{2}\right)^2,
\label{3.A.7} 
\end{eqnarray}
and the parameters $\beta_{i}$ are defined as
\begin{eqnarray}
\beta_{1}&=&\frac{\mathcal{E}}{4\Omega}-\frac{1}{4\mathcal{E}\,\Omega}\left(\nu\sqrt{k^{2}+M^{2}}+\frac{S_{z}}{8}\right)^{2}-\frac{1}{4}-\frac{j}{2\alpha};\label{3.A.8}\\
\beta_{2}&=&\frac{\mathcal{E}}{4\Omega}-\frac{1}{4\mathcal{E}\,\Omega}\left(\nu\sqrt{k^{2}+M^{2}}+\frac{S_{z}}{8}\right)^{2}+\frac{1}{4}-\frac{j}{2\alpha}.
\label{3.A.9} 
\end{eqnarray} 

Let us perform a change of variables  by the rule $\xi=\mathcal{E}\,\Omega\,r^{2}$, and thus, write Eq. (\ref{3.A.6}) as
\begin{eqnarray}
\xi\frac{d^{2}u_{i}}{d\xi^{2}} + \frac{du_{i}}{d\xi}+\left[\beta_{i}-\frac{\xi}{4}-\frac{\zeta^{2}_{i}}{4\xi}\right]u_{i}=0, 
\label{3.A.10}
\end{eqnarray}
where the solution to Eq. (\ref{3.A.10}) is given in the form:
\begin{eqnarray}
u_{i}(\xi)=e^{-\frac{\xi}{2}}\,\xi^{\frac{\left|\zeta_{i}\right|}{2}}\,F_i\left(\xi\right),
\label{3.A.11}
\end{eqnarray}
where $F_i\left(\xi\right)$ is the solution to the following equation:
\begin{eqnarray}
\xi\frac{d^{2}F_i}{d\xi^{2}}+\left(\left|\zeta_{i}\right|+1-\xi\right)\,\frac{dF_i}{d\xi}+\left(\beta_{i}-\frac{1}{2}-\frac{\left|\zeta_{i}\right|}{2}\right)F_i=0,
\label{3.A.11a}
\end{eqnarray}
which is  the confluent hypergeometric differential equation \cite{arf}, and the function $F_i\left(\xi\right)=\,_{1}F_{1}\left(-\left[\beta_{i}-\frac{1}{2}-\frac{\left|\zeta_{i}\right|}{2}\right],\,\left|\zeta_{i}\right|+1;\,\xi\right)$ is the confluent hypergeometric function.  It is well-known that the confluent hypergeometric  function becomes a polynomial of degree $n$  if
\begin{eqnarray}
n=\beta_{i}-\frac{1}{2}-\frac{\left|\zeta_{i}\right|}{2},
\label{3.A.12}
\end{eqnarray} 
where $n=0,1,2,3,\ldots$, then, from Eq. (\ref{3.A.12}), we obtain the allowed energy levels for both spinors
\begin{eqnarray}
\mathcal{E}_{n,\,j}&=&2\Omega\left[n+\frac{|j|}{2\alpha}+\frac{j}{2\alpha}+\frac{1}{2}\right]\nonumber\\
[-2mm]\label{3.A.13.a}\\[-2mm]
&\pm& \sqrt{ 4\Omega^{2}\left[n+\frac{|j|}{2\alpha}+\frac{j}{2\alpha}+\frac{1}{2}\right]^{2}+\left(\nu\sqrt{k^{2}+M^{2}}+\frac{S^{z}}{8}\right)^{2}},\nonumber
\end{eqnarray} 
where $n=0,1,2,3,\ldots$ and $j$ is half-integer. Hence, Eq. (\ref{3.A.13.a}) corresponds to the allowed energies of a Dirac particle in the Som-Raychaudhuri space-time with a cosmic string. In the case $\alpha=1$, we recover the eigenvalues of energy for a Dirac particle in the Som-Raychaudhuri space-time discussed in Ref. \cite{figueiredo}. Besides, the energy levels  (\ref{3.A.13.a}) are analogous to the relativistic Landau levels for fermions since they have an infinite degeneracy for $\alpha=1$. In the presence of the cosmic string ($\alpha\neq 1$), therefore, the degeneracy of the energy levels {\bf is} broken.   Furthermore, the presence of the torsion term, $ \frac{S^{z}}{8}$, shifts the eigenvalues (\ref{3.A.13.a}) in contrast to the case without torsion. Its presence splits each energy level in a doublet. This physical effect has been reported for case $\alpha=1$ in Ref. \cite{figueiredo}. We can also observe that this splitting produced by the torsion is additive  due to the existence of the constant of motion associated with the operator $\hat{C}$ given by first equation in (\ref{3.3}). Note in that the presence of torsion introduces an asymmetry in the energy levels when we consider $k=M=0$, i.e., the allowed energies become
\begin{eqnarray}
\bar{\mathcal{E}}_{n,\,m}&=& 2\Omega\left[n+\frac{|j|}{2\alpha}+\frac{j}{2\alpha}+\frac{1}{2}\right]\nonumber\\
[-2mm]\label{3.A.13b}\\[-2mm]
&\pm& \sqrt{4\Omega^{2}\left[n+\frac{|j|}{2\alpha} +\frac{j}{2\alpha} +\frac{1}{2}\right]^{2}+\left(\frac{S^{z}}{8}\right)^{2}}.\nonumber
\end{eqnarray}  
 Note that, we have in Eq. (\ref{3.A.13b}) the limit of the  Riemannian flatness with  a topological defect, i.e., the Minkowski  space-time metric with a topological defect plus a torsion field given by term proportional to $S^{z}$. The term $\left(\frac{S^{z}}{8\Omega}\right)^{2}$ is the contribution of the torsion to the eigenvalues. From Eq. (\ref{2.A.7}) we have $S^{z}=-2\Omega$, which yields $\left(\frac{S^{z}}{8}\right)^{2}=\frac{\Omega^{2}}{16}$. On the other hand, in the case of  zero torsion, we have that the contribution of the term $\left(\frac{S^{z}}{8}\right)^{2}$ is  zero, therefore, the allowed energies are analogous to the relativistic Landau levels for a Dirac particle in the cosmic string space-time \cite{cflandau1}, where the parameter associated with rotation plays the role of the cyclotron frequency, i.e., the rotation plays the role of the uniform magnetic field in the $z$-direction. In this way, the presence of term $S^{z}$ causes an asymmetry in the energy levels which are no longer purely Landau-type energy levels,  differently from the results obtained in Ref. \cite{figueiredo}.

Finally, the wave function for the Dirac particle in the Som-Raychaudhuri space-time in the presence of torsion with the cosmic string is given by
\begin{eqnarray}
\psi\left(t,\,\xi,\,\phi,\,z\right)&=&C_{n, j}\,e^{-\frac{\xi}{2}}e^{-i\left[\mathcal{E}t-j\phi-kz\right]}\times\nonumber\\
[-2mm]\label{3.A.14.a}\\[-2mm]
&\times&\left(\begin{array}{c}
\xi^{\frac{j}{\alpha}-\frac{1}{2}}\,_{1}F_{1}\left(-n,\,\frac{j}{\alpha}+\frac{1}{2};\,\xi\right)\\
\xi^{\frac{j}{\alpha} +\frac{1}{2}}\,_{1}F_{1}\left(-n,\,\frac{j}{\alpha}+\frac{3}{2};\,\xi\right)\\
\tau_{1}\,\xi^{\frac{j}{\alpha}-\frac{1}{2}}\,_{1}F_{1}\left(-n,\,\frac{j}{\alpha}+\frac{1}{2};\,\xi\right)\\
\tau_{2}\,\xi^{\frac{j}{\alpha} +\frac{1}{2}}\,_{1}F_{1}\left(-n,\,\frac{j}{\alpha}+\frac{3}{2};\,\xi\right)\\
 \end{array}\right),\nonumber
\end{eqnarray}
where $C_{n, j}$ is a constant and $j\geqslant\frac{1}{2}$. The solution for $j \leqslant -\frac{1}{2}$ is,
\begin{eqnarray}
\psi\left(t,\,\xi,\,\phi,\,z\right)&=&C_{n, j}\,e^{-\frac{\xi}{2}}e^{-i\left[\mathcal{E}t-j\phi-kz\right]}\times\nonumber\\
[-2mm]\label{3.A.14.b}\\[-2mm]
&\times&\left(\begin{array}{c}
\xi^{-\frac{j}{\alpha}+\frac{1}{2}}\,_{1}F_{1}\left(-n,\,-\frac{j}{\alpha}+\frac{3}{2};\,\xi\right)\\
\xi^{-\frac{j}{\alpha} -\frac{1}{2}}\,_{1}F_{1}\left(-n,\,-\frac{j}{\alpha}+\frac{1}{2};\,\xi\right)\\
\tau_{1}\,\xi^{-\frac{j}{\alpha}+\frac{1}{2}}\,_{1}F_{1}\left(-n,\,-\frac{j}{\alpha}+\frac{3}{2};\,\xi\right)\\
\tau_{2}\,\xi^{-\frac{j}{\alpha} -\frac{1}{2}}\,_{1}F_{1}\left(-n,\,-\frac{j}{\alpha}+\frac{1}{2};\,\xi\right)\\
 \end{array}\right),\nonumber
\end{eqnarray}
here $C_{n, j}$ is a constant as well.

\subsection{Spherically symmetric G\"odel-type space-time with a topological defect}\label{subsec5}

Let us consider the   spherically symmetric G\"odel space-time with a cosmic string in $z$-axis. This case corresponds to taking $l^{2}<0$ in the line element (\ref{geoplain}). Then, we wish to solve the Dirac equation (\ref{2.B.6}), where the corresponding operators $\hat{\pi}_{r}$ and $\hat{\pi}_{\phi}$ are given by
 \begin{eqnarray}
\hat{\pi}_{r}&=&-i\left[\left(1+\frac{r^{2}}{4R^{2}}\right)\frac{\partial}{\partial r}+\left(1-\frac{r^{2}}{4R^{2}}\right)\frac{1}{2r}\right];\nonumber\\
[-2mm]\label{3.B.1}\\[-2mm]
\hat{\pi}_{\phi}&=&-i\left(1+\frac{r^{2}}{4R^{2}}\right)\left[\frac{1}{\alpha r}\frac{\partial}{\partial\phi}- \frac{\Omega\,r}{\left(1+r^{2}/4R^{2}\right)}\frac{\partial}{\partial t}\right].\nonumber 
\end{eqnarray}

Thereby, by substituting the {\it ansatz} (\ref{3.4}) into the Dirac equation (\ref{2.B.6}), we have a set of four coupled differential equations: 
\begin{eqnarray}
\left[\mathcal{E}-\nu\sqrt{k^{2}+M^{2}}-\frac{S^{z}}{8}\right]u_{1}&=&-i\tau_{2}\left[\left(1+\frac{r^{2}}{4R^{2}}\right)\left(\frac{\partial}{\partial r}+\frac{j}{\alpha\,r}\right)+\left(1-\frac{r^{2}}{4R^{2}}\right)\frac{1}{2r}+\mathcal{E}\,\Omega\, r\right]u_{2};
\nonumber\\
\left[\mathcal{E}+\nu\sqrt{k^{2}+M^{2}}+\frac{S^{z}}{8}\right]u_{2}&=&-i\tau_{1}\left[\left(1+\frac{r^{2}}{4R^{2}}\right)\left(\frac{\partial}{\partial r}-\frac{j}{\alpha r}\right)+\left(1-\frac{r^{2}}{4R^{2}}\right)\frac{1}{2r}-\mathcal{E}\,\Omega\,r\right]u_{1};
\nonumber\\
\label{3b3}\\
\tau_{1}\left[\mathcal{E}-\nu\sqrt{k^{2}+M^{2}}-\frac{S^{z}}{8}\right]u_{1}&=&-i\left[\left(1+\frac{r^{2}}{4R^{2}}\right)\left(\frac{\partial}{\partial r}+\frac{j}{\alpha r}\right)+\left(1-\frac{r^{2}}{4R^{2}}\right)\frac{1}{2r}+\mathcal{E}\,\Omega\,r\right]u_{2};
\nonumber\\
\tau_{2}\left[\mathcal{E}+\nu\sqrt{k^{2}+M^{2}}+\frac{S^{z}}{8}\right]u_{2}&=&-i\left[\left(1+\frac{r^{2}}{4R^{2}}\right)\left(\frac{\partial}{\partial r}-\frac{j}{\alpha r}\right)+\left(1-\frac{r^{2}}{4R^{2}}\right)\frac{1}{2r}-\mathcal{E}\,\Omega\, r\right]u_{1}.
\nonumber
\end{eqnarray}

From the coupled equations given in (\ref{3b3}), it is possible to  obtain two second order differential equations for each spinor given by
\begin{eqnarray}
\left(1+\frac{r^{2}}{4R^{2}}\right)^{2}\left[\frac{d^{2}u_{i}}{dr^{2}}+\frac{1}{r}\frac{d u_{i}}{d r}\right]-\left[a'_{i}\,r^{2}+\frac{b^{2}_{i}}{r^{2}}-c_{i} \right]u_{i}=0,
\label{3.B.7}
\end{eqnarray}
where $i=1,2$ denotes each spinor. The parameters $a'_{i}$ are defined as
\begin{eqnarray}
a'_{1}&=&\frac{a^{2}_{1}}{16R^{4}}=\frac{1}{16R^{4}}\left(\frac{j}{\alpha}+\frac{1}{2}+4R^{2}\,\mathcal{E}\,\Omega\right)^{2};\nonumber\\
[-2mm]\label{3.B.8}\\[-2mm]
a'_{2}&=&\frac{a^{2}_{2}}{16R^{4}}=\frac{1}{16R^{4}}\left(\frac{j}{\alpha}-\frac{1}{2}+4R^{2}\mathcal{E}\,\Omega\right)^{2},\nonumber
\end{eqnarray}
the parameters $b_{i}$ are defined in the form:
\begin{eqnarray}
b^{2}_{1}&=&\left(\frac{j}{\alpha}-\frac{1}{2}\right)^{2};\nonumber\\
[-2mm]\label{3.B.10}\\[-2mm]
b^{2}_{2}&=& \left(\frac{j}{\alpha}+\frac{1}{2}\right)^{2},\nonumber
\label{3.B.11}
\end{eqnarray}
and finally, the parameters $c_{i}$ are
\begin{eqnarray}
c_{1}&=&\mathcal{E}^{2}-\left(\nu\sqrt{k^{2}+M^{2}}+\frac{S^{z}}{8}\right)^{2}-\frac{3}{8R^{2}}-\mathcal{E}\,\Omega-\frac{j^{2}}{2\alpha^{2}\,R^{2}}-\frac{2j\mathcal{E}\,\Omega}{\alpha};\nonumber\\
[-2mm]\label{3.B.12}\\[-2mm]
c_{2}&=&\mathcal{E}^{2}-\left(\nu\sqrt{k^{2}+M^{2}}+\frac{S^{z}}{8}\right)^{2}-\frac{3}{8R^{2}}+\mathcal{E}\,\Omega-\frac{j^{2}}{2\alpha^{2}\,R^{2}}-\frac{2j\mathcal{E}\,\Omega}{\alpha}.\nonumber
\end{eqnarray}
Now, let us introduce a new variable through the relation: $r=2R\tan\theta$; thus, Eq. (\ref{3.B.7}) becomes:
\begin{eqnarray}
\frac{d^2 u_{i}}{d \theta^{2}} +\left(\frac{1}{\cos\theta\,\sin\theta}-\frac{2\sin\theta}{\cos\theta}\right)\frac{d u_{i}}{d\theta}-\left[a^{2}_{i}\frac{\sin^{2}\theta}{\cos^{2}\theta}+b^{2}_{i}\frac{\cos^{2}\theta}{\sin^{2}\theta}-4R^{2} c_{i}\right]u_{i}=0.
\label{3.B.13}
\end{eqnarray}
We can go further by defining $x=\cos\theta$ and then we can write $\xi=1-x^{2}$, therefore, Eq. (\ref{3.B.13}) is rewritten in the form:
\begin{eqnarray}
\xi\left(1 -\xi\right)\frac{d^{2} u_{i}}{d \xi^{2}} +\left(1-2\xi\right)\frac{d u_{i}}{d\xi}-\left[\frac{a^{2}_{i}}{4}\frac{\xi}{\left(1-\xi\right)}+\frac{b^{2}_{i}}{4}\frac{\left(1-\xi\right)}{\xi}- R^{2}\,c_{i}\right]u_{i} = 0.
\label{3.B.14}
\end{eqnarray}

By labeling $\lambda_i=\frac{\left|a_{i}\right|}{2}$ and $\delta_i=\frac{\left|b_{i}\right|}{2}$, we can write the solutions to Eq. (\ref{3.B.14}) as
\begin{eqnarray}
u_{i}\left(\xi\right)=\xi^{\delta_i}\left(1-\xi\right)^{\lambda_i}\,\bar{F}_{i}\left(\xi\right), 
\label{3.B.15}
\end{eqnarray}
where $\bar{F}_{i}\left(\xi\right)$ are unknown functions. By substituting Eq. (\ref{3.B.15}) into Eq. (\ref{3.B.14}), we obtain two non-coupled equation for $\bar{F}_{1}\left(x\right)$ and $\bar{F}_{2}\left(x\right)$, 
\begin{eqnarray}
\xi\left(1-\xi\right)\frac{d^{2} \bar{F}_{i}}{d \xi^{2}}+\left[ 2\delta_i + 1 - 2\xi\left(\delta_i + \lambda_i + 1\right)\right]\frac{d \bar{F}_{i}}{d\xi}
 -\left[\lambda_{i} + 2\lambda_{i}\delta_{i} + \delta_{i} - R^{2}c_i\right] \bar{F}_{i}=0,
\label{3.B.16}
\end{eqnarray} 

In Eq. (\ref{3.B.16}), we have two decoupled differential equations which correspond to the hypergeometric differential equations \cite{arf,abra} and the functions $\bar{F}_{i}\left(x\right)=\,_{2}F_{1}\left(A,\,B,\,C_i;\,\xi\right)$ are the hypergeometric functions, where the coefficients $A$ and $B$ are given by
\begin{eqnarray}
A&=& \left[\left(\frac{j}{\alpha} + \frac{1}{2}\right) + 2R^{2}\mathcal{E}\Omega\right] + \sqrt{4R^{4}\Omega^{2}\mathcal{E}^{2} +  R^{2}\left[\mathcal{E}^{2} - \left(\nu\sqrt{k^{2}+M^{2}} + \frac{S_z}{8}\right)^2\right]};\nonumber\\
\label{3.17}\\
B&=& \left[\left(\frac{j}{\alpha} + \frac{1}{2}\right) + 2R^{2}\mathcal{E}\Omega\right] - \sqrt{4R^{4}\Omega^{2}\mathcal{E}^{2} + R^{2}\left[\mathcal{E}^{2} - \left(\nu\sqrt{k^{2}+M^{2}} + \frac{S_z}{8}\right)^2\right]}.\nonumber
\end{eqnarray}

Next, by imposing that the hypergeometric series becomes a polynomial of degree $n$, then, we obtain (for both spinors)
\begin{eqnarray}\label{sphere} 
\mathcal{E}_{n,\,j}&=& 2\Omega\left[n+\frac{|j|}{2\alpha}+\frac{j}{2\alpha} +\frac{1}{2}\right]\nonumber\\
&\pm& 2\Omega\left\{\left[n+\frac{|j|}{2\alpha}+ \frac{j}{2\alpha}+\frac{1}{2}\right]^{2}+\frac{1}{4R^{2}\,\Omega^{2}}\left[n+\frac{|j|}{2\alpha}+\frac{j}{2\alpha}+\frac{1}{2}\right]^{2}+\right. \nonumber \\
&+&\left. \frac{1}{4\Omega^{2}}\left(\nu\sqrt{k^{2}+M^{2}}+\frac{S^{z}}{8}\right)^{2}\right\}^{1/2},
\end{eqnarray}
where $n=0,1,2,3,\ldots$ and $-n +\frac{1}{2}\leqslant \frac{j}{\alpha}\leqslant 4\Omega R^2\mathcal{E} + \frac{1}{2}$.  

Hence, Eq. (\ref{sphere})  gives the allowed energies for a Dirac particle in the spherical G\"odel-type space-time with a cosmic string. In the case $\alpha=1$, we recover the relativistic energy levels for a Dirac particle in the spherical G\"odel space-time obtained in Ref. \cite{figueiredo}.  Note that the case where the eigenvalue $\mathcal{E}$ has a finite degeneracy was observed in Refs. \cite{figueiredo,fiol,josevi},  and in Ref. \cite{dunne} for Landau levels in a spherical space. This finite degeneracy has lower and upper bounds defined by $-n +\frac{1}{2}\leq \frac{j}{\alpha} \leq 4\Omega\,R^{2}\,\mathcal{E}+\frac{1}{2}$. Furthermore, if $\alpha=1$, the finite degeneracy is given by $-n +\frac{1}{2}\leq j \leq 4\Omega\,R^{2}\,\mathcal{E}+\frac{1}{2}$. Therefore, when $\alpha\neq 1$, the degeneracy of the relativistic energy levels is broken due to the presence of the cosmic string. It is worth pointing out that the presence of torsion introduces a modification in the zero point of energy when we consider the case $k=M=0$. This fact is well known. In the case $R\rightarrow\infty$, we obtain the same results obtained in the previous subsection (for Som-Raychaudhuri space-time). The presence of the torsion term $\frac{S^{z}}{8}$ in Eq. (\ref{sphere}) also splits each energy level in a doublet. This physical effect has been observed in previous subsection for the Som-Raychaudhuri case and also for the case $\alpha=1$ in Ref. \cite{figueiredo}. Moreover, we observe that this splitting given by torsion is additive due to the existence of a constant of motion associated with $\hat{C}$ (see Eq. (\ref{3.3})). In this way, we have the influence of the curvature, the topology of the defect and the torsion of this family of spherical G\"odel space-time on energy levels (\ref{sphere}).      

The corresponding wave function for the spherical case of the G\"odel-type solution with a cosmic string is given by
\begin{eqnarray}
\psi(t,\,\xi,\,\phi,\, z)&=&\bar{C}_{n, m}\,e^{-i\left[\mathcal{E}t- j\phi-kz\right]}\times\\\nonumber
&\times&\left(\begin{array}{c}
\left(1-\xi\right)^{\frac{\left|a_{1}\right|}{2}}\,\xi^{\frac{\left|b_{1}\right|}{2}}\,_{2}F_{1}\left(A,\,B,\,\frac{j}{\alpha}+\frac{1}{2};\,\xi\right) \\
\left(1-\xi\right)^{\frac{\left|a_{2}\right|}{2}}\,\xi^{\frac{\left|b_{2}\right|}{2}}\,_{2}F_{1}\left(A,\,B,\,\frac{j}{\alpha}+\frac{3}{2};\,\xi\right)\\ 
\tau_{1}\left(1-\xi\right)^{\frac{\left|a_{1}\right|}{2}}\,\xi^{\frac{\left|b_{1}\right|}{2}}\,_{2}F_{1}\left(A,\,B,\,\frac{j}{\alpha} +\frac{1}{2};\,\xi\right)\\
\tau_{2}\left(1-\xi\right)^{\frac{\left|a_{2}\right|}{2}}\,\xi^{\frac{\left|b_{2}\right|}{2}}\,_{2}F_{1}\left(A,\,B,\frac{j}{\alpha} +\frac{3}{2};\,\xi\right)
 \end{array}\right),
\label{3.B.20}
\end{eqnarray}
where $\bar{C}_{n, m}$ is a constant. The parameters $A$ and $B$ of the hypergeometric functions have been defined in Eq. (\ref{3.17}) and $\frac{1}{2}\leqslant \frac{j}{\alpha}\leqslant 4\Omega R^2\mathcal{E} + \frac{1}{2}$. We also have
\begin{eqnarray}
\psi(t,\,\xi,\,\phi,\, z)&=&\bar{C}_{n, m}\,e^{-i\left[\mathcal{E}t- j\phi-kz\right]}\times\\\nonumber
&\times&\left(\begin{array}{c}
\left(1-\xi\right)^{\frac{\left|a_{1}\right|}{2}}\,\xi^{\frac{\left|b_{1}\right|}{2}}\,_{2}F_{1}\left(1-A,\,1-B,\,-\frac{j}{\alpha}+\frac{3}{2};\,\xi\right) \\
\left(1-\xi\right)^{\frac{\left|a_{2}\right|}{2}}\,\xi^{\frac{\left|b_{2}\right|}{2}}\,_{2}F_{1}\left(1-A,\,1-B,\,-\frac{j}{\alpha}+\frac{1}{2};\,\xi\right)\\ 
\tau_{1}\left(1-\xi\right)^{\frac{\left|a_{1}\right|}{2}}\,\xi^{\frac{\left|b_{1}\right|}{2}}\,_{2}F_{1}\left(1-A,\,1-B,\,-\frac{j}{\alpha} +\frac{3}{2};\,\xi\right)\\
\tau_{2}\left(1-\xi\right)^{\frac{\left|a_{2}\right|}{2}}\,\xi^{\frac{\left|b_{2}\right|}{2}}\,_{2}F_{1}\left(1-A,\,1-B,-\frac{j}{\alpha} +\frac{1}{2};\,\xi\right)
 \end{array}\right),
\label{3.B.20a}
\end{eqnarray}
where $\bar{C}_{n, m}$ is a constant and $-n+\frac{1}{2}\leqslant \frac{j}{\alpha}\leqslant -\frac{1}{2}$.

\subsection{Hyperbolic G\"odel-type space-time with a topological defect}\label{subsec6}

Henceforth, we consider hyperbolic G\"odel-type space-time with a cosmic string  corresponding to the case $l^{2}>0$ in the line element (\ref{geoplain}). Our focus is on the search for  bound-state solutions to the Dirac equation (\ref{2.C.6}). In contrast to the previous cases, the operators $\hat{\pi}_{r}$ and $\hat{\pi}_{\phi}$ have the following form: 
\begin{eqnarray}
\hat{\pi}_{r}&=&-i\left[\left(1-l^{2}\,r^{2}\right)\frac{\partial}{\partial r} + \left(1+l^{2}\,r^{2}\right)\frac{1}{2r}\right];\nonumber\\
[-2mm]\label{3.C.1}\\[-2mm]
\hat{\pi}_{\phi}&=&-i\left(1-l^{2}\,r^{2}\right)\left[\frac{1}{\alpha r}\frac{\partial}{\partial\phi} + \frac{\Omega r}{\left(1-l^{2}\,r^{2}\right)}\frac{\partial}{\partial t}\right],\nonumber 
\label{3.C.2}
\end{eqnarray}
and, by using the {\it ansatz} (\ref{3.4}) into the Dirac equation (\ref{2.C.6}), we have a new set of four coupled differential equations: 
\begin{eqnarray}
\left[\mathcal{E}-\left(\nu\sqrt{k^{2}+M^{2}}+\frac{S^{z}}{8}\right)\right]u_{1}&=&-i\tau_{2}\left[\left(1-l^{2}\,r^{2}\right)\left(\frac{\partial}{\partial r}+\frac{j}{\alpha r}\right)+\left(1+l^{2}\,r^{2}\right)\frac{1}{2r}+\mathcal{E}\,\Omega\, r\right]u_{2};
\nonumber\\
\left[\mathcal{E}+\left(\nu\sqrt{k^{2}+M^{2}}+\frac{S^{z}}{8}\right)\right]u_{2}&=&-i\tau_{1}\left[\left(1-l^{2}\,r^{2}\right)\left(\frac{\partial}{\partial r}-\frac{j}{\alpha r}\right)+\left(1+l^{2}\,r^{2}\right)\frac{1}{2r}-\mathcal{E}\,\Omega\,r\right]u_{1};
\nonumber\\
\label{3c3}\\
\tau_{1}\left[\mathcal{E}-\left(\nu\sqrt{k^{2}+M^{2}}+\frac{S^{z}}{8}\right)\right]u_{1}&=&-i\left[\left(1-l^{2}\,r^{2}\right)\left(\frac{\partial}{\partial r}+\frac{j}{\alpha r}\right)+\left(1+l^{2}\,r^{2}\right)\frac{1}{2r}+\mathcal{E}\,\Omega\,r\right]u_{2};
\nonumber\\
\tau_{2}\left[\mathcal{E}+\left(\nu\sqrt{k^{2}+M^{2}}+\frac{S^{z}}{8}\right)\right]u_{2}&=&-i\left[\left(1-l^{2}\,r^{2}\right)\left(\frac{\partial}{\partial r}-\frac{j}{\alpha r}\right)+\left(1+l^{2}\,r^{2}\right)\frac{1}{2r}-\mathcal{E}\,\Omega\,r\right]u_{1}.\nonumber
\end{eqnarray}

From Eqs. (\ref{3c3}), we obtain two non-coupled equations which can be written as
\begin{eqnarray}
\left(l^{2}\,r^{2}-1\right)^{2}\left[\frac{d^2 u_{i}}{d r^{2}}+\frac{1}{r}\frac{d u_{i}}{d r}\right]- \left[a'_{i}\,r^{2}+\frac{b^{2}_{i}}{r^{2}}-c_{i}\right]u_{i}=0,
\label{3.C.7}
\end{eqnarray}
where $i =1,\,2$ denotes which spinor. The parameters $a'_{i}$ are
\begin{eqnarray}
a'_{1}&=& l^{4}\,a^{2}_{1}=l^{4}\left(\frac{j}{\alpha}+\frac{1}{2}-\frac{\mathcal{E}\,\Omega}{l^{2}}\right)^{2};\nonumber\\
[-2mm]\label{3.C.8}\\[-2mm] 
a'_{2}&=& l^{4}\,a^{2}_{2}=l^{4}\left(\frac{j}{\alpha}-\frac{1}{2}-\frac{\mathcal{E}\,\Omega}{l^{2}}\right)^{2},\nonumber
\end{eqnarray}
while the parameters $b_{i}$ are
\begin{eqnarray}
b^{2}_{1}&=&\left(\frac{j}{\alpha}-\frac{1}{2}\right)^{2};\nonumber\\
[-2mm]\label{3.C.10}\\[-2mm]
b^{2}_{2}&=&\left(\frac{j}{\alpha}+\frac{1}{2}\right)^{2},\nonumber
\end{eqnarray}
and, finally, the parameters $c_{i}$ are defined as
\begin{eqnarray}
c_{1}&=&\mathcal{E}^{2}-\left(\nu\sqrt{k^{2}+M^{2}}+\frac{S^{z}}{8}\right)^{2}+\frac{3l^{2}}{2}-\mathcal{E}\,\Omega+\frac{2l^{2}\,j^{2}}{\alpha^{2}}- \frac{2j\mathcal{E}\,\Omega}{\alpha};\nonumber\\
[-2mm]\label{3.C.12}\\[-2mm]
c_{2}&=&\mathcal{E}^{2}-\left(\nu\sqrt{k^{2}+M^{2}}+\frac{S^{z}}{8}\right)^{2}+\frac{3l^{2}}{2}+\mathcal{E}\,\Omega+\frac{2l^{2}\,j^{2}}{\alpha^{2}}-\frac{2j\mathcal{E}\,\Omega}{\alpha}.\nonumber
\end{eqnarray}

We proceed with a change of variables given by $r = \dfrac{\tanh(l\theta)}{l}$, then, Eq. (\ref{3.C.7}) becomes
\begin{eqnarray}
\frac{d^2 u_{i}}{d \theta^{2}}&+&\left(\frac{2l\sinh\left(l\theta\right)}{\cosh\left(l\theta\right)}+ \frac{l}{\cosh\left(l\theta\right)\sinh\left(l\theta\right)}\right)\frac{d u_{i}}{d\theta}\nonumber\\
&-&\left[a^{2}_{i}l^{2}\frac{\sinh^{2}\left(l\theta\right)}{\cosh^{2}\left(l\theta\right)}+b^{2}_{i}\,l^{2}\,\frac{\cosh^{2}\left(l\theta\right)}{\sinh^{2}\left(l\theta\right)}-c_{i}\right]u_{i}=0,
\label{3.C.13}
\end{eqnarray}
and by defining $y=\cosh(l\theta)$ and $\xi=y^{2}- 1$, we obtain 
\begin{eqnarray}
\xi\left(1+\xi\right)\frac{d^2 u_{i}}{d \xi^{2}} +\left(1+2\xi\right)\frac{d u_{i}}{d\xi}-\left[\frac{a^{2}_{i}}{4}\frac{\xi}{\left(1+\xi\right)}+\frac{b^{2}_{i}}{4}\,\frac{\left(1+\xi\right)}{\xi}-\frac{\varepsilon^{2}-1}{4}\right]u_{i}=0,
\label{3.C.14}
\end{eqnarray}
with $\frac{\varepsilon^{2}-1}{4}= \frac{c_{i}}{4l^{2}}$. Let us call $\lambda_i =\frac{\left|a_{i}\right|}{2}$ and $\delta_i =\frac{\left|b_{i}\right|}{2}$, thus, the solutions to Eq. (\ref{3.C.14}) are given in the form:
\begin{eqnarray}
u_{i}\left(\xi\right)=\xi^{\delta_i}\,\left(1+\xi\right)^{\lambda_i}\,\tilde{F}_{i}\left(\xi\right),
\label{3.C.15} 
\end{eqnarray}
where $\tilde{F}_{i}\left(\xi\right)$ are unknown functions. After some calculations, we obtain the following equation for $\tilde{F}_{i}\left(x\right)$:
\begin{eqnarray}
\xi\left(1 -\xi\right)\frac{d^{2} \tilde{F}_{i}}{d \xi^{2}}+\left[2\delta_i + 1 -2\xi\left(\delta_i + \lambda_i + 1\right)\right]\frac{d \tilde{F}_{i}}{d\xi} -\left[\lambda_i + 2\lambda_i\delta_i +\delta_i +\frac{c_i}{4l^{2}}\right]\tilde{F}_{i}=0,
\label{3.C.16}
\end{eqnarray}

Again, we obtain two second order differential equations that correspond to the hypergeometric differential equations \cite{arf,abra}, where the functions $\tilde{F}_{i}\left(\xi\right)=\,_{2}F_{1}\left(A,\,B,\,C;\,\xi\right)$ are the hypergeometric functions. The parameters $A$ and $B$ are
\begin{eqnarray}
A&=&\left[\left(\frac{j}{\alpha} + \frac{1}{2}\right) - \frac{\mathcal{E}\Omega}{2l^{2}}\right]+ \sqrt{\frac{\mathcal{E}^{2}\Omega^{2}}{4l^{4}} - \frac{1}{4l^{2}}\left[\mathcal{E}^{2} - \left(\nu\sqrt{k^{2}+M^{2}} + \frac{S^z}{8}\right)^2\right]}\nonumber\\
\label{3.C.18}\\
B&=& \left[\left(\frac{j}{\alpha} + \frac{1}{2}\right) - \frac{\mathcal{E}\Omega}{2l^{2}}\right]- \sqrt{\frac{\mathcal{E}^{2}\Omega^{2}}{4l^{4}} - \frac{1}{4l^{2}}\left[\mathcal{E}^{2} - \left(\nu\sqrt{k^{2}+M^{2}} + \frac{S^z}{8}\right)^2\right]}\nonumber
\end{eqnarray}

As in the previous subsection, we impose that the hypergeometric series becomes a polynomial of degree $n$, then, we obtain the following allowed energies for both spinors:
\begin{eqnarray}
\mathcal{E}_{n,\,j}&=& 2\Omega\left[n+\frac{|j|}{2\alpha}+ \frac{j}{2\alpha}+\frac{1}{2}\right]\nonumber\\
&\pm& 2\Omega\left\{\left[n+\frac{|j|}{2\alpha}+\frac{j}{2\alpha}+\frac{1}{2}\right]^{2}-\frac{l^{2}}{\Omega^{2}}\left[n+ \frac{|j|}{2\alpha}+\frac{j}{2\alpha} +\frac{1}{2}\right]^{2}\right. \nonumber\\
 &+&\left. \frac{1}{4\Omega^{2}}\left(\nu\sqrt{k^{2}+M^{2}}+\frac{S^{z}}{8}\right)^{2}\right\}^{1/2},
\label{3.C.19} 
\end{eqnarray}
where $n=0,1,2,3,\ldots$ and $-n+\frac{1}{2}\leqslant\frac{j}{\alpha}\leqslant \infty$. 

As  observed in previous works \cite{figueiredo, fiol, josevi}, the energy levels in the  hyperbolic geometry  have two contributions: one is  a discrete energy levels given in Eq. (\ref{3.C.19})  and a second contribution to energy levels that varies continuously, whose lower limit is also determined by the relation $-n+\frac{1}{2}\leqslant\frac{j}{\alpha}\leqslant -\frac{1}{2}$ . Thus, in the hyperbolic G\"odel-type space-time with a topological defect, the spectra of energy can be discrete or continuous, where they are determined  by the parameter $\varepsilon$ through  the expression: 
\begin{eqnarray}
\varepsilon^{2}=\frac{\left(\Omega^{2}-l^{2}\right)}{4l^{4}}\,\mathcal{E}^{2} + \frac{1}{4l^{2}}\left(\nu\sqrt{k^{2}+M^{2}}+\frac{S^{z}}{8}\right)^{2};
\label{3.C.20}
\end{eqnarray}
thus, we have discrete values of energy for $\varepsilon^{2}\,\geq\,1$ and a continuous spectrum of energy for $\varepsilon^{2}\,<\,1$. 

We go further by analyzing the relativistic energy levels (\ref{3.C.19}) through the parameters $\Omega$ and $l$. Then, we have:
\begin{enumerate}
	\item For $\Omega^{2}\,>\,l^{2}$, we obtain from Eq. (\ref{3.C.20}) a discrete spectrum of energy when
\begin{eqnarray}
\mathcal{E}^{2}&\geq&\frac{l^{2}}{\Omega^{2} - l^{2}}\left[4l^{2} -\left(\nu\sqrt{k^{2} + M^{2}} +\frac{S^{z}}{8}\right)^{2}\right].
\label{3.C.21}
\end{eqnarray}

Otherwise, the relativistic spectrum of energy is continuous.
	
	\item For $\Omega^{2}=l^{2}$, we obtain from Eq. (\ref{3.C.20}) a discrete spectrum of energy when
\begin{eqnarray}
\left(\nu\sqrt{k^{2} + M^{2}} +\frac{S^{z}}{8}\right)^{2}\geq 4l^{2}
\label{3.C.23}
\end{eqnarray}
Otherwise, the relativistic spectrum of energy is continuous.

    \item For $\Omega^{2}\,<\,l^{2}$, we also obtain from Eq. (\ref{3.C.20}) a discrete spectrum of energy when
\begin{eqnarray}
\mathcal{E}^{2}&\leq&\frac{l^{2}}{l^{2} -\Omega^{2}}\left[\left(\nu\sqrt{k^{2} + M^{2}} +\frac{S^{z}}{8}\right)^{2} - 4l^{2}\right].
\label{3.C.24}
\end{eqnarray}

Otherwise, we also have that the relativistic spectrum of energy is continuous.
\end{enumerate}

 Again, the presence of the torsion term $\frac{S^{z}}{8}$ in the eigenvalues (\ref{3.C.19}) splits the energy levels in a doublet. This splitting is also additive due to the existence of the same constant of motion associated with $\hat{C}$, as we can see in Eq. (\ref{3.3}). Therefore, we have the influence of curvature, torsion and the topology of the defect of this family of hyperbolic G\"odel space-time on the energy eigenvalues.     

Finally, the corresponding wave function for the hyperbolic case of the G\"odel-type solution with a cosmic string is given by
\begin{eqnarray}
\psi(t,\,\xi,\,\phi,\,z)&=&\tilde{C}_{n,\,m}\,\,e^{-i\left[\mathcal{E}t-j\phi-kz\right]}\times\nonumber\\
&\times&\left(\begin{array}{c}
\left(1+\xi\right)^{\frac{\left|a_{1}\right|}{2}}\,\xi^{\frac{\left|b_{1}\right|}{2}}\,_{2}F_{1}\left(A,\,B,\,\frac{j}{\alpha}+\frac{1}{2};\,\xi\right)\\
\left(1+\xi\right)^{\frac{\left|a_{2}\right|}{2}}\,\xi^{\frac{\left|b_{2}\right|}{2}}\,_{2}F_{1}\left(A,\,B,\,\frac{j}{\alpha}+\frac{3}{2};\,\xi\right)\\
\tau_{1}\left(1+\xi\right)^{\frac{\left|a_{1}\right|}{2}}\,\xi^{\frac{\left|b_{1}\right|}{2}}\,_{2}F_{1}\left(A,\,B,\,\frac{j}{\alpha}+\frac{1}{2};\,\xi\right)\\
\tau_{2}\left(1+\xi\right)^{\frac{\left|a_{2}\right|}{2}}\,\xi^{\frac{\left|b_{2}\right|}{2}}\,_{2}F_{1}\left(A,\,B,\,\frac{j}{\alpha} +\frac{3}{2}; \xi\right)
\end{array}\right),
\label{3.C.26}
\end{eqnarray}
where $\tilde{C}_{n,\,m}$ is a constant. The parameters $A$ and $B$ of the hypergeometric functions have been defined in Eq. (\ref{3.C.18}) and $\frac{1}{2}\leqslant\frac{j}{\alpha}\leqslant\infty$. We still have 
\begin{eqnarray}
\psi(t,\,\xi,\,\phi,\,z)&=&\tilde{C}_{n,\,m}\,\,e^{-i\left[\mathcal{E}t-j\phi-kz\right]}\times\nonumber\\
&\times&\left(\begin{array}{c}
\left(1+\xi\right)^{\frac{\left|a_{1}\right|}{2}}\,\xi^{\frac{\left|b_{1}\right|}{2}}\,_{2}F_{1}\left(1-A,\,1-B,\,-\frac{j}{\alpha}+\frac{3}{2};\,\xi\right)\\
\left(1+\xi\right)^{\frac{\left|a_{2}\right|}{2}}\,\xi^{\frac{\left|b_{2}\right|}{2}}\,_{2}F_{1}\left(1-A,\,1-B,\,-\frac{j}{\alpha}+\frac{1}{2};\,\xi\right)\\
\tau_{1}\left(1+\xi\right)^{\frac{\left|a_{1}\right|}{2}}\,\xi^{\frac{\left|b_{1}\right|}{2}}\,_{2}F_{1}\left(1-A,\,1-B,\,-\frac{j}{\alpha}+\frac{3}{2};\,\xi\right)\\
\tau_{2}\left(1+\xi\right)^{\frac{\left|a_{2}\right|}{2}}\,\xi^{\frac{\left|b_{2}\right|}{2}}\,_{2}F_{1}\left(1-A,\,1-B,\,-\frac{j}{\alpha} +\frac{1}{2}; \xi\right)
\end{array}\right),
\label{3.C.26a}
\end{eqnarray}
with $\tilde{C}_{n,\,m}$ as constant and $-n +\frac{1}{2}\leqslant\frac{j}{\alpha}\leqslant-\frac{1}{2}$.

Hence, we have obtained in Eq. (\ref{3.C.19}) the allowed energies for a Dirac particle in the hyperbolic G\"odel-type space-time with a cosmic string, where we have analyzed the conditions for achieving a discrete or a continuous spectrum of energy. It is worth observing that for $\alpha=1$, therefore,  we recover in Eq. (\ref{3.C.19}) the relativistic energy levels for a Dirac particle in the hyperbolic G\"odel-type space-time \cite{figueiredo}, where they are infinitely degenerate even though bounds exist. Due to this infinitely degeneracy, the relativistic energy levels (\ref{3.C.19}) for $\alpha=1$ are analogous to the Landau levels for fermions in hyperbolic space \cite{comtet,dunne}. On the other hand, with $\alpha\neq 1$ in Eq. (\ref{3.C.19}), we have that the degeneracy of the relativistic energy levels are broken due to influence of the cosmic string.  Besides, the presence of torsion in the space-time gives rise to a contribution to the energy levels  (\ref{3.C.19}) even if in the case $k=M=0$. In the case $R\rightarrow \infty$  or $l\rightarrow 0$, we recover the results obtained in the Som-Raychaudhuri space-time. Note that the allowed energies (\ref{3.C.19}) can be discrete or continuous, however, in both cases the presence of torsion in the space-time modifies the ground state energy of system.

\section{Conclusions}\label{sec4}

In this contribution, we have investigated the behaviour of a Dirac particle in a class of G\"odel  background space-times  with a topological defect in the  Einstein-Cartan theory. We have obtained the corresponding Dirac equations in the Som-Raychaudhury, spherical and hyperbolic  background space-times  with torsion that contain a cosmic string that passes through $z$-axis, and then, solved them analytically. In the case of the Som-Raychaudhury space-time with a cosmic string, we have obtained the allowed energies for this relativistic quantum system and shown an analogy between the relativistic energy levels and the Landau levels, where the rotation plays the role of the uniform magnetic field in the $z$-direction. We have also seen that the presence of the topological defect breaks the degeneracy of the relativistic energy levels.

In the case of the  spherically symmetric G\"odel-type space-time with a cosmic string, we have also obtained the corresponding Dirac equation and solved it analytically. We have obtained the allowed energies of the relativistic system. We have seen that degeneracy of the energy levels has lower and upper bounds that decrease with the presence of the topological defect. Again, we  found an analogy between the allowed energies and the Landau levels, where the rotation plays the role of the uniform magnetic field in the $z$-direction. The presence of torsion has also modified the  ground state  of the energy. We have also obtained and solved analytically the Dirac equation in the hyperbolic G\"odel-type space-time with a cosmic string. We have shown that the allowed energies for the system can be discrete or continuous. Moreover, the energy levels are analogous to the Landau levels in a hyperbolic space, where the presence of the topological defect breaks the degeneracy of the relativistic energy levels. We have also observed that presence of torsion in the space-time modifies the relativistic energies in both discrete and continuous cases.

  In all cases, we have seen that there exist the  contribution of the torsion term $\frac{S^{z}}{8}$ in the allowed energies (see Eqs. (\ref{3.A.13b}) ,(\ref{sphere}) and (\ref{3.C.19})). The torsion effect on the allowed energies corresponds to the splitting of each energy level in a doublet. Besides, the torsion contribution is additive due to the constant of motion associated with the operator $\hat{C}$ given in Eq. (\ref{3.3}). Hence, we have observed the influence of curvature, torsion and the topology of the defect of the families of flat, spherical and hyperbolic G\"odel  background space-times pierced by a topological defect in the energy eigenvalues.      

 In a recent work, Drukker and Fiol \cite{fiol} have dealt with scalar particles in a G\"odel-type space-time  by investigating the possible holographic description of a single chronological safe region. They discussed the use of this results in Hall droplets of finite size \cite{halp}.  We claim that the present study can be used to investigate the Hall effect in curved space $S^{3}$ \cite{nair,nair2}. Due to the analogy with the Landau quantization, the present study could be used to investigate the higher dimensional quantum Hall effects and the A-class topological insulators with emphasis on the noncommutative geometry \cite{hasebe}. Other interests are in studies of the quantum Hall effect in the fuzzy sphere \cite{fuzzy,hu,fuzzy2}. Based on the present results, we can apply these results in the quantum Hall effect in fuzzy sphere \cite{fuzzy} by making use of the ideas given in Ref. \cite{hu}. Furthermore, in Ref. \cite{fuzzy}, the Landau levels of a Dirac particle are studied in the hyperbolic space and the fuzzy sphere \cite{fuzzy2}. In this way, the results of Refs. \cite{fuzzy,fuzzy2} can be connected with the present work with the purpose of investigating the Hall effect in this geometry. Finally, we claim that the tools employed in this article to investigate the quantum dynamics of fermions in the families of the G\"odel-type solutions with torsion can be used in systems of condensed matter physics where both curvature and torsion are important.
 
\acknowledgments{The authors would like to thank the Brazilian agencies CNPq, CAPES and FAPESQ for financial support. We are grateful to Albert Petrov  for critical reading of this manuscript  }

\end{document}